# Driver Assistant: Persuading Drivers to Adjust Secondary Tasks Using Large Language Models

Wei Xiang[1], Muchen Li[2], Jie Yan[3], Manling Zheng[4], Hanfei Zhu[5], Mengyun Jiang[6], and Lingyun Sun*

*Abstract*— Level 3 automated driving systems allows drivers to engage in secondary tasks while diminishing their perception of risk. In the event of an emergency necessitating driver intervention, the system will alert the driver with a limited window for reaction and imposing a substantial cognitive burden. To address this challenge, this study employs a Large Language Model (LLM) to assist drivers in maintaining an appropriate attention on road conditions through a "humanized" persuasive advice. Our tool leverages the road conditions encountered by Level 3 systems as triggers, proactively steering driver behavior via both visual and auditory routes. Empirical study indicates that our tool is effective in sustaining driver attention with reduced cognitive load and coordinating secondary tasks with takeover behavior. Our work provides insights into the potential of using LLMs to support drivers during multi-task automated driving.

## I. INTRODUCTION

Level 3 automated driving systems allow drivers to perform secondary tasks while driving, yet drivers still need to pay attention to the road conditions. However, when drivers become immersed in secondary tasks, they often become distracted. When sudden road hazards appear or the system requires manual takeover, drivers need to rapidly switch from a relaxed state to a highly alert state. This abrupt transition creates significant cognitive load, potentially leading to safety risks. Therefore, the key challenge lies in how to maintain an appropriate level of driver awareness of the environment while allowing them to perform secondary tasks.

Persuasion is a method of changing behavior and supporting safe driving practices [1]. Compared to alerts, persuasion might establish comfortable and gentle communication with the driver [2], [3], [4]. Research attempts to apply persuasion to assist drivers in changing behaviors. Unlike warnings to highlight risks [5], [6], while hazards have not occurred to persuade during automated driving remains a challenge. It is unclear when LLM should persuade the driver and how persuasion methods should be applied.

This study proposes an LLM-based persuasion tool and explores the impact of persuasion on driver behaviors. Employing LLM's information processing and natural language processing capabilities, this tool supports decisions as a copilot and changes drivers' behavior by persuasive methods. We first conducted a semi-structured interview to understand drivers' preferences for persuasion. Then, we developed an LLM-based tool as "driving assistant" to maintain appropriate alertness while multitasking. An empirical study involving 40 participants validated the effectiveness of our tool.

## II. BACKGROUND

### A. Risk Factors During Automated Driving

Secondary tasks in automated driving may not be seen as potential risks but rather are considered part of the daily commute [7], [8]. Research indicates that drivers in vehicles are frequently distracted by using smartphones, eating, and adjusting in-vehicle devices [9], [10]. Distractions will change drivers' risk perception and endanger road safety [11]. Mobile phone usage negatively impacts driving performance [12]. Adjusting in-vehicle devices, such as radios, DVD players and navigation also significantly interferes with driving [13]. Distractions in the vehicle can impact drivers' performance and readiness to respond to emergencies [14].

Road conditions affect driving behavior and risk [15]. Risk factors like vehicle and road conditions significantly impact traffic safety [16-17]. In urban scenario, factors such as signage, road conditions, visibility, vehicles, traffic signals, pedestrians, and obstacles (such as animals or roadside trees) are the primary determinants of varying risk levels [18]. Based on the annual accident reports in China, H. Chen et al. analyzed the primary risk factors of traffic accidents and proposed the city scenarios as the basis for assessing road risks [19]. Factors such as traffic volume, pedestrians, road conditions, lighting, and weather are considered the main influences in assessing road risks.

Promptly helping drivers respond to potential emergencies is a big challenge. The unavoidable secondary tasks divert drivers' attention during automated driving. They reduce the driver's control over the vehicle and increase the likelihood of traffic accidents in automated driving [20]. Balancing drivers' attention between secondary tasks and road conditions is still a key challenge for driving systems.

### B. Intervention that Changes Driving Behavior

Unlike warning cues, persuasion can establish a long-term collaboration with drivers and strategically improve their behavior to enhance road safety. Currently, based on Fogg Behavior Model (FBM), persuasion guidance promotes behavior change through improving executive ability, reminding unreasonable behaviors and increasing behavioral motivation [21]. Meanwhile, Fogg identified five elements of human-computer interaction: physical, psychological, linguistic, social dynamics, and social roles. This has facilitated the practical application of persuasive strategies in human-computer interaction. The Persuasive Systems Design (PSD), a persuasive system framework based on FBM theory, providing guidance for the construction of persuasive system within human-computer interaction [22]. In addition, "Nudge" is a strategy that guides expected behavior changes through minor adjustments. Caraban et al. introduced 23 nudging mechanisms, grouped into six categories to help individuals make better decisions [23].

Interaction methods such as voice prompts and interface interactions in automated driving systems have been widely adopted and proven to be effective mechanisms for guiding behavior change in persuasive systems [24]. However, risks in the situation are dynamic. These studies focus on offering drivers immediate risk responses but overlook how interaction improves their behavior. Such interaction can prevent or avert critical events earlier in L3 automated driving.

*C. LLM Application in Driving and Persuasion*

Research demonstrates that LLM help enhance driving safety. Sha et al. proved that LLM can serve as an effective decision-maker in complex scenarios, showing notable safety, efficiency, universality, and interactivity within autonomous driving systems. Jiageng Mao et al. discovered that LLM, acting as a driver assistant, has intuitive common sense and robust reasoning capabilities [24], enabling more detailed and humanized autonomous driving. Relevant studies indicate that LLMs possess the capability for risk identification [25], which can be utilized in the judgment and assessment of driving risks.

LLM has also been verified to be suitable for implementing various persuasion applications [26]. Harrison et al. utilized LLM to tailor persuasive plans for different users by combining images and information [27]. Remountakis et al. improved the hotel service recommendation system by integrating LLM and persuasive strategy and explored the potential of persuasion in communication [28]. These instances show that LLM effectively processes information and generates context-specific persuasive content.

To adapt to driving scenarios, the persuasion provided by LLM should consider the criticality of the driving environment, drivers' level of distraction, and offer reliable and preferable advice.

## III. FORMATIVE STUDY

This section conducted semi-structured user interviews to explore drivers' perceptions of secondary tasks and persuasive preference in L3 automated driving. A total of 20 drivers were interviewed, lasting about 20 minutes. The sample had a balanced gender distribution (11 males and 9 females) and an average driving experience of 9 years (SD = 7.9).

*A. Driver Distraction*

All drivers agreed that frequent secondary tasks divert attention and pose a threat to road safety. Drivers expressed their views on the degree of distraction caused by secondary tasks and generally acknowledged that these behaviors are difficult to balance in L3 automated driving. We highlighted several representative opinions:

"Attention needs to be more concentrated in areas with many pedestrians and vehicles. Conversely, with fewer pedestrians and clear visibility, moderate distraction is acceptable" (P15). Additionally, drivers shared that they felt considerable pressure when they received an emergency takeover request from the L3 automated driving system, which made them more aware of safety risks while driving and added to their psychological burden. They also showed reluctance to be abruptly awakened and requested to take over while driving.

Subsequently, we compiled the drivers' ratings for typical secondary tasks: eating (3.8), reaching for objects (3.9), using a mobile phone (6.7), and adjusting in-vehicle devices (4.7). We reviewed the research on the risk probabilities of various secondary tasks and found that they largely align with the feedback from the scales [29]. Based on drivers' assessments of secondary tasks, we consider them as a reference for LLM's distraction risk evaluation in experimental design.

*B. Drivers' Persuasion Preferences*

Drivers are attentive to the interactive modes of persuasion and exhibit a high level of receptiveness towards guided actions. It was observed that, compared to drivers with less experience, those with more experience prefer brief and relaxed communication with driving assistants. Drivers expressed willingness to accept persuasion from the driving assistant, but dislike frequent interruptions. They also had great expectations regarding the mode of interaction.

Based on drivers' feedback, we summarized their preferences and suggestions for persuasive strategies, taking into account specific expectations for a driving assistant. To enable LLM to better understand and generate appropriate persuasive content, we have outlined four key principles of persuasion: 1. Keep it simple and crisp. 2. Provide direct, reliable advice without resorting to commands or demands. 3. Avoid emphasizing bad consequences. 4. Be colloquial, avoiding seriousness, like people's everyday conversations.

## IV. TOOL DESIGN

The tool includes two functions: First, assessing road risks and drivers' attention to determine the necessity of persuasion; Second, generating persuasion content. Based on Fogg's persuasion model and the framework provided by PSD, the tool develops six persuasive strategies. Persuasive system employs GPT-4-0613's API. We identified the trigger timing of the LLM in the system based on drivers' assessments of secondary tasks and road risks. Additionally, we collected drivers' preferences to guide the LLM in generating persuasive suggestions and appropriately presenting them. The overall flowchart is shown in Figure 1.

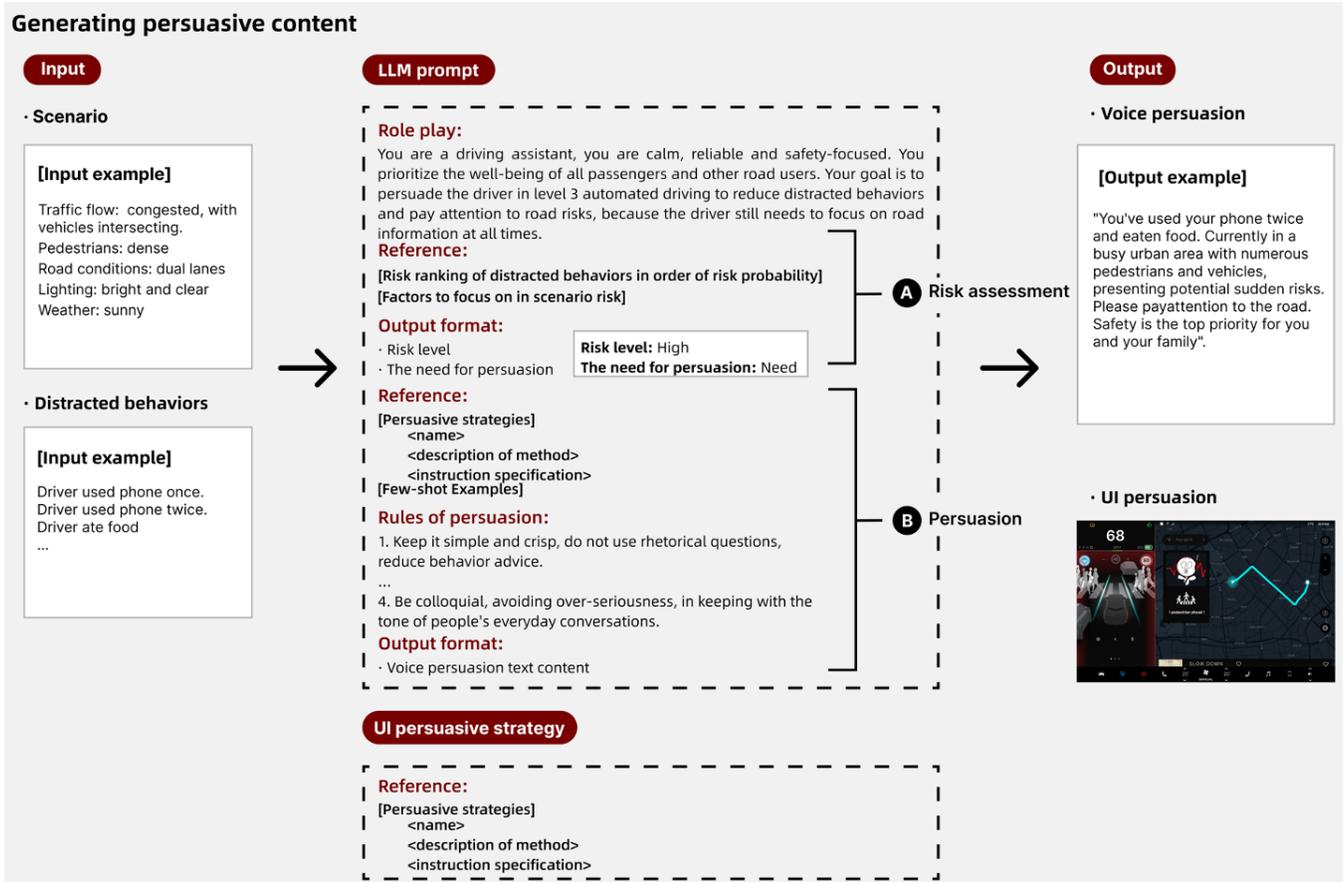

Figure 1. A Flowchart for Generating Persuasive Content.

## A. The Criteria for Timing of Persuasion

The system determines when to trigger persuasive interventions by evaluating two key factors: road risk information and driver attention allocation. For road risk assessment, we use five parameters to describe road risk: traffic flow, pedestrian presence and movement, road conditions, lighting, and weather conditions, which have been collected in existing driving systems and input for LLM. Distracted driving behaviors are identified by tracking the driver's eye movements using Tobii Glasses 3 eye tracker. Four typical distracting behaviors are used to determine persuasion conditions based on the level of attention allocation : using a smartphone, adjusting in-car devices, drinking water, and reaching for an item (from high to low). The driver's distracted behavior and frequency within a specific period (e.g., 30 seconds) were monitored. Both road risk data and driver behavior are converted to timestamped text format and processed by the LLM, supporting the decision of persuasion.

## B. The Timing and Condition for Persuasion

We evaluated how consistently LLM and human drivers judge when persuasion is needed using the Kappa metric. Three participants with L3 automated driving expertise and 5+ years of driving experience assessed 30 LLM-generated scenarios. These scenarios, combining various road conditions and secondary tasks, were labeled as either "require persuasion" or "do not require persuasion". The kappa values between the LLM and the participants were 0.661, 0.664, and 0.724, while the kappa values between the participants were 0.795, 0.789, and 0.865. These findings indicate a high level of agreement among drivers and LLM when evaluating the persuasion trigger of generated scenarios.

## C. Persuasion Advice Generation

Our persuasive system follows six persuasive strategies: 1. Status Feedback; 2. Emphasize Risk; 3. Default Concern (making choices for the driver); 4. Reliable Advice; 5. Social Connection; 6. Social Interactions. The presentations and descriptions of these strategies are shown in TABLE I.

TABLE I. PERSUASION STRATEGIES AND DESCRIPTION.

| Element | Strategy | Description of Method |
| --- | --- | --- |
| Remind unreasonable behaviors | Status Feedback | Timely reminders and environmental hazard feedback to attract attention (UI). |
| | Emphasize Risk | Enhance driver alertness and improve response time between action and readiness. |
| Improve execution ability | Default Concern | Simplify task steps to facilitate decision-making. |
| | Reliable Advice | Guide the driver to focus on risky matters based on the scenario and driver condition. |
| Increase behavioral motivation | Social Connection | Establish connections, evoke a sense of communication to maintain safety jointly. |
| | Social Interaction | Emotional expression, intuitive reflection; Give encouragement and reward. |

## D. Persuasive UI Design

User interface design is inspired by Tesla's in-vehicle interface, as shown in Figure 2. The system added a cartoon character. This avatar reflects the driver's current status through different expressions. When the driver is focused and in good condition, the avatar appears lively and happy. Conversely, the avatar looks tense and anxious. Additionally, when the driver shows great driving behavior, the assistant provides positive feedback with "encouragement".

The screen displays basic information. As the risk increases, the interface border will change to yellow/red to alert the driver that the risk is rapidly increasing. The interface will also display changes in pedestrian and vehicle flow, road conditions, and other traffic information.

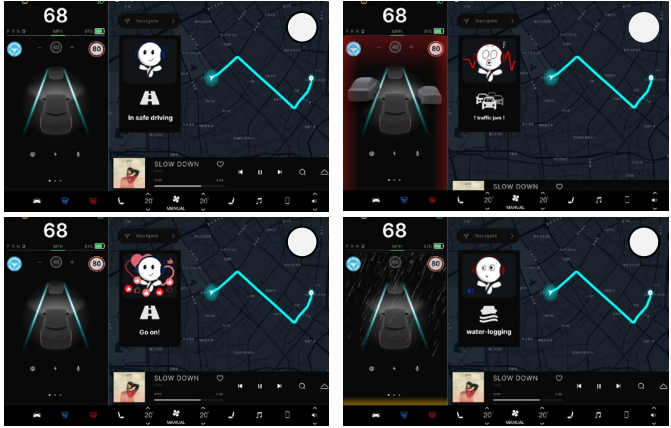

Figure 2. Interface. (a) Default situation. (b) Road congestion at a high-risk level. (c) Risk-free situation, the avatar provides positive feedback. (d) Raining condition at a low-risk level, the yellow border flickers. The text in the UI is translated from Chinese.

## V. EXPERIMENT

The experiment aims to evaluate the impact of the persuasion system on drivers. The experiment set up a simulated driving cockpit and used Unity 3D to create simulated driving scenes.

### A. Participant

The experiment involved 40 participants. Participants must meet the following criteria: Have at least 3 years of driving experience.

TABLE II. SCENARIO RISK SECTIONS

| Risk | Risk Factors |
|---|---|
| No risk | Clear weather, good visibility, dual lanes with light traffic, and few pedestrians on the road. |
| Low risk | The weather was moderate rain, the road surface was slippery, and the light was gloomy. |
| Medium risk | Construction section, only in a single lane of traffic, cloudy weather, limited visibility. |
| High risk | In the evening, visibility is poor, heavy traffic, and vehicle congestion. |

The final analysis was conducted with an effective sample size of 36. There were 21 males and 15 females, with an average age of 25.81 years (SD=7.837) and an average driving experience of 4.41 years (SD=3.97). After the experiment, participants will receive compensation of 50 RMB.

### B. Risk Sections of the Road

The experiment studied the impact of different road risk sections on driver behavior. Based on situational risk factors [24], the experiment includes four sections: no risk, low risk, medium risk, and high risk. Each section includes different risk factors, as shown in TABLE Ⅱ.

Drivers will acquire road risk factors based on a first-person driving perspective, which accurately simulates the real driving experience. The driving perspective includes the steering wheel, the road ahead, and the rearview mirrors.

### C. Hardware Equipment

To replicate an authentic driving environment as closely as possible, the cockpit was outfitted with the following devices: (1) a driver's seat; (2) a steering wheel; (3) a windshield; (4) an eye tracker; and (5) an in-car interface, as shown in Figure 3.

### D. Baseline

We compare our proposed system against conventional takeover request audio alerts, which serve as the experimental baseline. The baseline system aims to redirect driver attention from secondary tasks back to road monitoring during automated driving. It provides drivers with verbal messages when encountering weather, traffic, and events [30].

Based on our experimental scenario design, we configured four warning messages about road environmental conditions: "Raining outside", "Increased pedestrian activity ahead", "Construction zone approaching", and "Entering urban area". These alerts are triggered automatically based on predefined road conditions without considering the driver's current attention state or distraction level.

### E. Procedure

Participants are required to complete two phases: the baseline system experiment (20 minutes) and the persuasion system experiment (20 minutes). The vehicle will traverse four different risk level sections, each lasting 5 minutes. During the experiment, participants are allowed to perform the following distracted tasks: reading tweets on a smartphone, typing and sending "Hello, 2025" on the in-car device, retrieving an item from the back seat, and drinking water.

Throughout both experimental phases, we collected objective measures included the number of secondary tasks and eye movement data captured via Tobii Glasses 3. Following each experimental phase, participants completed subjective evaluations through questionnaires and interviews to provide feedback on systems and share their suggestions.

### F. Subjective interview

For subjective assessments, we used a cognitive load and experience questionnaire based on related research. The experience questionnaire is divided into three parts: Perceived Usefulness (PU), Perceived Ease of Use (PEOU), and Behavioral Intention (BI). Participants were asked to assign subjective scores ranging from 1 to 5 to various dimensions related to both systems. The questionnaire includes: Positive Influence (Q1); Driving Improvement (Q2); Accuracy (Q3); Contribution to Safe Driving (Q4); Distraction Adjustment (Q5); Comprehensibility (Q6); Experience (Q7); Comfort (Q8); Not Causing Disturbance (Q9); Not Causing

Distractions (Q10); Willingness to Change (Q11); Willingness to Use (Q12).

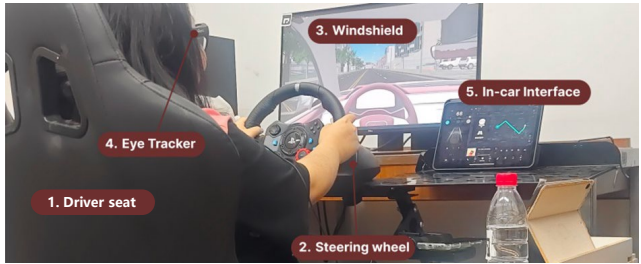

Figure 3. Experiment in progress: the participant sits in a simulated cockpit.

## VI. RESULTS

### A. The Number of Secondary Tasks

The average number of secondary tasks and standard deviations were presented in TABLE Ⅲ. The results revealed significant differences in the frequency of secondary tasks of the subjects during automated driving in the two experimental sessions. According to descriptive statistics, participants engaged in significantly fewer secondary tasks when exposed to the persuasive system compared to the baseline (F = 6.40, p < 0.05, η² = 0.17). Under varying risk levels, a noticeable increase in secondary tasks occurred as the risk level decreased (F=25.52, p<0.001, η²=0.44). Additionally, the interaction between the "baseline vs. persuasion system" (bp) and "risk level" also had a significant impact. In the baseline condition, changes in risk level exerted a greater influence on secondary tasks (F=10.70, p<0.001, η²=0.25).

TABLE III. STATISTICS OF THE NUMBER OF SECONDARY TASKS

| Risk Level | Baseline | | Persuasion | |
|---|---|---|---|---|
| | *Mean* | *SD* | *Mean* | *SD* |
| High risk | 1.66 | 1.305 | 3.11 | 1.891 |
| Medium risk | 5.20 | 3.104 | 3.80 | 2.336 |
| Low risk | 5.46 | 2.944 | 3.66 | 2.209 |
| No risk | 6.31 | 4.164 | 4.51 | 2.994 |

### B. Eye Movement

#### 1) The Duration of Whole Fixation

The eye-tracking data of drivers during secondary tasks were collected and analyzed. Results showed a significant difference between the persuasive system and the baseline system in the duration of whole fixation (F = 5.955, p = 0.022, η² = 0.192), and a significant interaction effect was found between the systems and road risk levels (F = 8.939, p < 0.001, η² = 0.263). The relationship between the systems and road risk levels showed a linear trend (F = 13.989, p = 0.001, η² = 0.359). The persuasive system exhibited a generally shorter duration of whole fixation and maintained a more consistent distribution across driving. The baseline system had a higher duration of whole fixation in the low-risk condition, which decreased significantly under medium and high risk.

#### 2) Drivers' Pupil Diameter

The results indicated a significant difference in the average pupil diameter between the systems, along with a significant interaction effect with road risk levels (F = 5.209, p = 0.015, η² = 0.172). A linear relationship was also found between the average pupil diameter and road risk under different systems (F = 6.532, p = 0.017, η² = 0.207). In high-risk conditions, the average pupil diameter showed more significant changes in the persuasive system compared to the baseline system.

### C. Subjective Evaluation

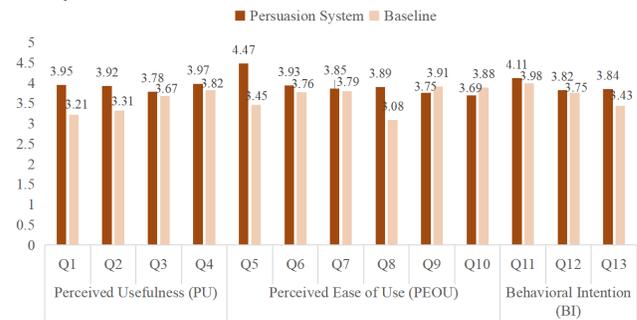

Figure 4. Subjective evaluation questionnaire results.

The results presented in Figure 4 demonstrate that the persuasion system outperforms the baseline across PU and BI metrics, while also noticeably scoring higher on comprehensibility and distraction management. These findings suggest that participants perceive the persuasion-based approach as more valuable for enhancing driving performance and safety, and expressing stronger intentions to adopt such a system.

## VII. DISCUSSION

### A. The Impact of Drivers' Performance in Persuasion System

Our study explored the impact of persuasive technology on improving driver behavior under automated driving conditions. The results indicated that, compared to warning alerts, drivers performed fewer secondary tasks under the persuasion approach. The frequency of secondary tasks tended to be balanced under low, medium, and high road risk conditions.

We explored the impact of persuasion on drivers' physiological indicators. We observed that the change in average pupil diameter for the persuasive system was smaller compared to the baseline system. This suggests that under the persuasive system, drivers experienced relatively lower cognitive load. The fixation duration in the persuasion system is more evenly distributed across different risk sections, indicating a more stable attention allocation by drivers in the persuasion system. Therefore, we believe the persuasion system supports reasonable attention allocation and is more effective in influencing driver behavior with medium and low-risk conditions.

### B. Driver Attitude of Persuasion System

The evaluation results indicated that drivers generally found the persuasion system more understandable than warning alerts and positively impacted minimizing distraction and improving behavior. Most drivers felt that the persuasion system provided a good experience. They appreciated its impact on reducing distractions without feeling lectured.

### C. Limitation

The scenario is a virtual scene built using Unity and experimented with by simulating a real environment through a driving position. Applications in real scenarios should be considered in subsequent research. Experimental results indicate that drivers also pay attention to personalized settings such as persuasion tone and role. These elements will be incorporated into future persuasive systems.

Additionally, further research need a more larger sample size for accurate results. Future also need more work to validate the practical applications of our persuasive system.

VIII. CONCLUSION

In this paper, we explore a method that utilizes LLM to persuade drivers other than warning alerts. The experimental results show that the LLM can effectively understand the persuasion and generate content that meets the user expectations, helping drivers to regulate devotion to secondary tasks. This suggests that in the context of human-machine collaborative driving, the assistant may evolve to become more of a partner, establishing a closer relationship with the driver while ensuring road safety.

REFERENCES

[1] J. K. Caird, C. R. Willness, P. Steel, and C. Scialfa, "A meta-analysis of the effects of cell phones on driver performance," *Accid. Anal. Prev.*, vol. 40, no. 4, pp. 1282–1293, Jul. 2008, doi: 10.1016/j.aap.2008.01.009.
[2] S. G. Charlton, "Driving while conversing: Cell phones that distract and passengers who react," *Accid. Anal. Prev.*, vol. 41, no. 1, pp. 160–173, Jan. 2009, doi: 10.1016/j.aap.2008.10.006.
[3] B. De Vos, A. Cuenen, V. Ross, H. Dirix, K. Brijs, and T. Brijs, "The Effectiveness of an Intelligent Speed Assistance System with Real-Time Speeding Interventions for Truck Drivers: A Belgian Simulator Study," *Sustainability*, vol. 15, no. 6, p. 5226, Mar. 2023, doi: 10.3390/su15065226.
[4] T. A. Dingus, V. L. Neale, S. G. Klauer, A. D. Petersen, and R. J. Carroll, "The development of a naturalistic data collection system to perform critical incident analysis: An investigation of safety and fatigue issues in long-haul trucking," *Accid. Anal. Prev.*, vol. 38, no. 6, pp. 1127–1136, Nov. 2006, doi: 10.1016/j.aap.2006.05.001.
[5] K. Duan, X. Yan, X. Li, and J. Hang, "Improving drivers' merging performance in work zone using an in-vehicle audio warning," *Transp. Res. Part F Traffic Psychol. Behav.*, vol. 95, pp. 297–321, May 2023, doi: 10.1016/j.trf.2023.04.004.
[6] G. M. Fitch et al., "The Impact of Hand-Held and Hands-Free Cell Phone Use on Driving Performance and Safety-Critical Event Risk," Art. no. DOT HS 811 757, Apr. 2013, Accessed: Apr. 02, 2025.
[7] S. P. McEvoy et al., "Role of mobile phones in motor vehicle crashes resulting in hospital attendance: a case-crossover study," *BMJ*, vol. 331, no. 7514, p. 428, Aug. 2005, doi: 10.1136/bmj.38537.397512.55.
[8] T. Teodorovicz, A. L. Kun, R. Sadun, and O. Shaer, "Multitasking while driving: A time use study of commuting knowledge workers to assess current and future uses," *Int. J. Hum.-Comput. Stud.*, vol. 162, p. 102789, Jun. 2022, doi: 10.1016/j.ijhcs.2022.102789.
[9] T. Horberry, J. Anderson, M. A. Regan, T. J. Triggs, and J. Brown, "Driver distraction: The effects of concurrent in-vehicle tasks, road environment complexity and age on driving performance," *Accid. Anal. Prev.*, vol. 38, no. 1, pp. 185–191, Jan. 2006, doi: 10.1016/j.aap.2005.09.007.
[10] J. Lee, T. Hirano, T. Hano, and M. Itoh, "Conversation during Partially Automated Driving: How Attention Arousal is Effective on Maintaining Situation Awareness," in *2019 IEEE International Conference on Systems, Man and Cybernetics (SMC)*, Oct. 2019, pp. 3718–3723. doi: 10.1109/SMC.2019.8914632.
[11] J. De Oña, R. De Oña, L. Eboli, C. Forciniti, and G. Mazzulla, "How to identify the key factors that affect driver perception of accident risk. A comparison between Italian and Spanish driver behavior," *Accid. Anal. Prev.*, vol. 73, pp. 225–235, Dec. 2014, doi: 10.1016/j.aap.2014.09.020.
[12] L. Luo, J. Ju, B. Xiong, Y.-F. Li, G. Haffari, and S. Pan, "ChatRule: Mining Logical Rules with Large Language Models for Knowledge Graph Reasoning," 2023, *arXiv*. doi: 10.48550/ARXIV.2309.01538.
[13] W. Li, K. Gkritza, and C. Albrecht, "The culture of distracted driving: Evidence from a public opinion survey in Iowa," *Transp. Res. Part F Traffic Psychol. Behav.*, vol. 26, pp. 337–347, Sep. 2014, doi: 10.1016/j.trf.2014.01.002.
[14] Y. Peng, L. N. Boyle, and J. D. Lee, "Reading, typing, and driving: How interactions with in-vehicle systems degrade driving performance," *Transp. Res. Part F Traffic Psychol. Behav.*, vol. 27, pp. 182–191, Nov. 2014, doi: 10.1016/j.trf.2014.06.001.
[15] N. Lyu, C. Deng, L. Xie, C. Wu, and Z. Duan, "A field operational test in China: Exploring the effect of an advanced driver assistance system on driving performance and braking behavior," *Transp. Res. Part F Traffic Psychol. Behav.*, vol. 65, pp. 730–747, Aug. 2019, doi: 10.1016/j.trf.2018.01.003.
[16] M. S. Young, J. M. Mahfoud, G. H. Walker, D. P. Jenkins, and N. A. Stanton, "Crash dieting: The effects of eating and drinking on driving performance," *Accid. Anal. Prev.*, vol. 40, no. 1, pp. 142–148, Jan. 2008, doi: 10.1016/j.aap.2007.04.012.
[17] H. Antonson, S. Mårdh, M. Wiklund, and G. Blomqvist, "Effect of surrounding landscape on driving behaviour: A driving simulator study," *J. Environ. Psychol.*, vol. 29, no. 4, pp. 493–502, Dec. 2009, doi: 10.1016/j.jenvp.2009.03.005.
[18] V. Beanland, A. J. Filtness, and R. Jeans, "Change detection in urban and rural driving scenes: Effects of target type and safety relevance on change blindness," *Accid. Anal. Prev.*, vol. 100, pp. 111–122, Mar. 2017, doi: 10.1016/j.aap.2017.01.011.
[19] H. Chen, Y. Zhao, and X. Ma, "Critical Factors Analysis of Severe Traffic Accidents Based on Bayesian Network in China," *J. Adv. Transp.*, vol. 2020, pp. 1–14, Nov. 2020, doi: 10.1155/2020/8878265.
[20] H. Singh and A. Kathuria, "Analyzing driver behavior under naturalistic driving conditions: A review," *Accid. Anal. Prev.*, vol. 150, p. 105908, Feb. 2021, doi: 10.1016/j.aap.2020.105908.
[21] B. J. Fogg, "Persuasive technology: using computers to change what we think and do," *Ubiquity*, vol. 2002, no. December, p. 2, Dec. 2002, doi: 10.1145/764008.763957.
[22] H. Oinas-Kukkonen and M. Harjumaa, "Persuasive Systems Design: Key Issues, Process Model, and System Features," *Commun. Assoc. Inf. Syst.*, vol. 24, 2009, doi: 10.17705/1CAIS.02428.
[23] A. Caraban, E. Karapanos, D. Gonçalves, and P. Campos, "23 Ways to Nudge: A Review of Technology-Mediated Nudging in Human-Computer Interaction," in *Proceedings of the 2019 CHI Conference on Human Factors in Computing Systems*, Glasgow Scotland Uk: ACM, May 2019, pp. 1–15. doi: 10.1145/3290605.3300733.
[24] S. Wang et al., "OmniDrive: A Holistic LLM-Agent Framework for Autonomous Driving with 3D Perception, Reasoning and Planning," 2024, *arXiv*. doi: 10.48550/ARXIV.2405.01533.
[25] Z. Zhou, H. Huang, B. Li, S. Zhao, Y. Mu, and J. Wang, "SafeDrive: Knowledge- and Data-Driven Risk-Sensitive Decision-Making for Autonomous Vehicles with Large Language Models," Dec. 19, 2024, *arXiv*: arXiv:2412.13238. doi: 10.48550/arXiv.2412.13238.
[26] J. Mao, Y. Qian, J. Ye, H. Zhao, and Y. Wang, "GPT-Driver: Learning to Drive with GPT," 2023, *arXiv*. doi: 10.48550/ARXIV.2310.01415.
[27] R. M. Harrison, A. Dereventsov, and A. Bibin, "Zero-Shot Recommendations with Pre-Trained Large Language Models for Multimodal Nudging," 2023, *arXiv*. doi: 10.48550/ARXIV.2309.01026.
[28] M. Remountakis, K. Kotis, B. Kourtzis, and G. E. Tsekouras, "ChatGPT and Persuasive Technologies for the Management and Delivery of Personalized Recommendations in Hotel Hospitality," 2023, *arXiv*. doi: 10.48550/ARXIV.2307.14298.
[29] "Visual-Manual NHTSA Driver Distraction Guidelines for In-Vehicle Electronic Devices," Federal Register. Accessed: Apr. 02, 2025.
[30] Y. Saito, Y. Watahiki, C. Leung, H. Zhou, and M. Itoh, "Effect of verbal messages with reminders to communicate driving situations to alter driver behavior in conditional driving automation," *Transp. Res. Part F Traffic Psychol. Behav.*, vol. 85, pp. 69–82, Feb. 2022, doi: 10.1016/j.trf.2022.01.003.